\documentclass[10pt, a4paper, twocolumn]{article}
\usepackage[utf8]{inputenc}
\usepackage[toc,page]{appendix}
\usepackage{amsmath}
\usepackage{amsfonts}
\usepackage{mathabx}
\usepackage{amssymb}
\usepackage[margin=.75in]{geometry}
\usepackage{color}
\usepackage{graphicx}
\usepackage{booktabs}
\usepackage{rotating}
\usepackage[font = small]{caption}
\usepackage{subcaption}
\usepackage{graphicx}
\usepackage{abstract}
\usepackage{url}
\usepackage{cite}
\usepackage{authblk}

\newcommand{\G}{\mathrm{G}}

\author[$\dagger$]{Simon R. Pocock}
\author[$\dagger$]{Paloma A. Huidobro}
\author[$\dagger$,$\ddagger$]{Vincenzo Giannini}

\affil[$\dagger$]{Physics Department, Blackett Laboratory, Imperial College London, Prince Consort Road, London SW7 2AZ}
\affil[$\ddagger$]{Instituto de Estructura de la Materia (IEM-CSIC), Consejo Superior de Investigaciones Cient{\'i}ficas, Serrano 121, 28006 Madrid, Spain}
\date{}

\title{Bulk-edge correspondence and long range hopping in the topological plasmonic chain}

\begin{document}
\twocolumn[
  \maketitle
  \begin{onecolabstract}
  The existence of topologically protected edge modes is often cited as a highly desirable trait of topological insulators. However, these edge states are not always present. A realistic physical treatment of long range hopping in a one-dimensional dipolar system can break the symmetry that protects the edge modes without affecting the bulk topological number, leading to a breakdown in bulk-edge correspondence. It is important to find a better understanding of where and how this occurs, as well as how to measure it. Here we examine the behaviour of the bulk and edge modes in a dimerised chain of metallic nanoparticles and in a simpler non-Hermitian next-nearest-neighbour model to provide some insight into the phenomena of bulk-edge breakdown. We construct bulk-edge correspondence phase diagrams for the simpler case and use these ideas to devise a measure of symmetry-breaking for the plasmonic system based on its bulk properties. This provides a parameter regime for which bulk-edge correspondence is preserved in the topological plasmonic chain, as well as a framework for assessing this phenomenon in other systems.
  \end{onecolabstract}
]

\section{Introduction}

Topological insulators (TIs) are often described as materials which have insulating bulks but support surface or edge states that are strongly protected from disorder or other perturbations by topology. Some time after the introduction of topological physics in the Hermitian quantum world~\cite{Thouless1982,Kane2005a,Kane2005b,Bernevig2006,Moore2010,Hasan2010}, photonic systems were shown to also exhibit topological properties~\cite{Haldane2008,Raghu2008,Wang2008,Wang2009,Khanikaev2012,Lu2014,Lu2016,Sun2017,Siroki2017,
Ozawa2018,Rider2019}. These photonic topological insulators (PTIs) have exciting applications for unidirectional waveguides~\cite{Wang2009} and lasing~\cite{St-Jean2017,Harari2018,Bandres2018} and show interesting effects in coupling to quantum emitters~\cite{Bello2018}. In addition, TIs have been shown to interact with light in intruiging ways~\cite{Siroki2016}. PTIs offer a useful platform to study how TIs are affected by non-Hermiticity, which can emerge as a consequence of loss, gain and phase information in photonic systems.

The topological properties of a system are specified by the symmetries of its Hamiltonian. In the Hermitian case a `periodic table’ of symmetry classes and the corresponding topological properties has been know for some time~\cite{Ryu2010}, while in comparison the non-Hermitian equivalent has been found very recently~\cite{Kawabata2018,Zhou2018}. Prior to this discovery some non-Hermitian symmetries had already been identified and studied in detail, such as the well know parity-time ($\mathcal{PT}$) symmetry, where in the photonic context loss and gain are carefully balanced so that the Hamiltonian has real valued eigenvalues~\cite{Ling2016,Feng2017,Lieu2018a}. In this work we consider a one-dimensional system with links to the famous Su-Schrieffer-Heeger (SSH) model~\cite{Su1979,Asboth2016}, whose topological properties emerge due to chiral symmetry. Chiral symmetry is often also referred to as sublattice symmetry, because the sublattices in a chirally symmetric system  are identical.

Topological insulators typically have a topological number associated with the bulk and surface or edge states depending on this number. Systems which are both in a topological phase and which have the correct number of surface or edge states related to the bulk topological number are said to have bulk-edge correspondence (BEC)~\cite{Rhim2018}, which is sometimes thought of as a requirement for a system to be a TI. There is, however, some discussion in the literature that this may be too narrow a definition and that some materials can be said to be topological in spite of a lack of edge or surface states~\cite{Fu2007,Hughes2011}. Even so, the existence of topologically protected edge modes is often desired. The question of the existence of BEC is a topic currently of great interest in the field of non-Hermitian TIs~\cite{Leykam2017,Xiong2018,Kunst2018,Jin2018,Alvarez2018,Zhong2018,Herviou2018,
Chen2019,Zirnstein2019,Song2019}, where some propose new topological numbers specific to non-Hermiticity~\cite{Zhong2018b}. Hermitian systems can also exhibit BEC breakdown due to breaking of the symmetry that protects the edge modes~\cite{PerezGonzalez2018}. 

In a previous work we studied the one-dimensional topological plasmonic chain with retardation and radiative effects and showed that it acts as a non-Hermitian topological insulator with edge modes due to an approximate chiral symmetry~\cite{Pocock2018}, as did others with a different mathematical approach~\cite{Downing2018}. In fact, when long range hopping is considered the chiral symmetry is ``trivially''~\cite{Pocock2018} broken by an identity term in the Hamiltonian, but the system still features the same topological numbers and phases because it retains inversion symmetry. Although technically any breaking of chiral symmetry removes topological protection, when the contribution from long range hopping is small enough edge modes still exist and feature some protection. However, in similar dipolar models such as cold atoms and phonon polaritons, for certain parameters of the chain the edge modes disappear while the Zak phase remains unchanged~\cite{Wang2018a,Wang2018b}. This is bulk-edge correspondence breakdown caused by chiral symmetry breaking due to long range hopping. Since realistic photonic systems often feature some degree of long range hopping it is necessary to understand how this affects the existence of bulk-edge correspondence. 

In this work we study BEC breakdown in this realistic physical model and also a simpler non-Hermitian SSH model, defining a model-specific measure of non-chirality to elucidate where BEC breakdown occurs.

\section{The topological plasmonic chain}

\begin{figure}[b]
\centering
\includegraphics[]{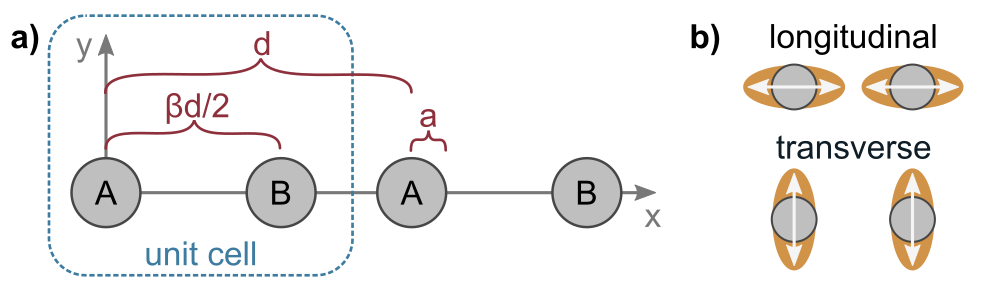}
\caption{(a) Schematic diagram of the topological plasmonic chain and (b) the two distinct and decoupled polarisations}
\label{fig:Fig0}
\end{figure}

The topological plasmonic chain in question is a one-dimensional chain of metallic nanoparticles with alternating spacing, depicted in figure~\ref{fig:Fig0}(a). There are two particles per unit cell labelled $A$ and $B$ with radius $a$ and unit cell spacing $d$. Intracell spacing is given by $\beta d/2$ so that the parameter $\beta$ describes the staggering of the chain, with $\beta=1$ equally spaced. The alternating spacing is reminiscent of the SSH model which has alternating hopping on a one-dimensional lattice~\cite{Su1979}, and reproduces the physics of the SSH model in the quasistatic limit~\cite{Ling2015,Downing2017,Gomez2017}. Similar SSH-like physics has also been studied in zigzag chains of nanoparticles~\cite{Slobozhanyuk2015,Kruk2018}.

When the centres of the particles are further apart than $3a$ the particles are well described using the coupled dipole approximation (CDA)~\cite{Park2004},

\begin{equation} \label{eqn:CDA}
\frac{1}{\alpha(\omega)}\mathbf{p}_n = \sum_{m\neq n}\G(\mathbf{r}_n,\mathbf{r}_m,\omega)\mathbf{p}_m,
\end{equation}

\noindent where $\mathbf{p}_n$ is the dipole moment of the $n$th particle, $\G(\mathbf{r}_n,\mathbf{r}_m,\omega)$ is the Green's dyadic between the positions $\mathbf{r}$ of the $n$th and $m$th particles at frequency $\omega$, $\alpha(\omega)$ the polarizability of the particles.

We use the modified long-wavelength approximation (MLWA) for the polarizability,

\begin{equation}
\alpha(\omega) = \frac{\alpha_{qs}(\omega)}{1-i\frac{2}{3}k^3\alpha_{qs}(\omega)-\frac{k^2}{a}\alpha_{qs}(\omega)},
\end{equation}

\noindent where the second term in the denominator accounts for radiative damping and the third is the dynamic depolarisation term~\cite{Jensen1999}. The wavevector magnitude is given by $k = \sqrt{\epsilon_B}\omega/c$, where $\epsilon_B$ is the background dielectric and $\alpha_{qs}(\omega)$ is the quasistatic polarizability given by

\begin{equation}
\alpha_{qs}(\omega) = a^3\frac{\epsilon(\omega)-\epsilon_B}{\epsilon(\omega)+2\epsilon_B}.
\end{equation}

\noindent Here $\epsilon(\omega)$ is the dielectric function of the particles, in this work given by the Drude model,

\begin{equation}
\epsilon(\omega) = \epsilon_\infty - \frac{\omega_P^2}{\omega^2+i\omega/\tau}.
\end{equation}

\noindent We consider silver nanoparticles in air, so that $\epsilon_\infty = 5$, $\hbar\omega_P = 8.9$\,eV, $\tau = 17$\,fs~\cite{Yang2015} and $\epsilon_B=1$. We take small particles with radius $a=5$\,nm, leading to a single particle surface plasmon resonance (SPR) frequency of $\hbar\omega_{sp} = 3.36$\,eV. These particles are small so that MLWA and quasistatic approximation for the polarizability give very similar results, but the use of MLWA allows the potential for larger particles. Typically the quasistatic approximation is reasonable below a radius of $20$\,nm and MLWA below $50$\,nm~\cite{Moroz2009}. The radius we consider here is large enough to treat the particles classically ($>2$\,nm)~\cite{Fitzgerald2016} but small enough to make the approximation that $\omega = \omega_{sp}$ in the Green's dyadic, linearising the function. This simplifies calculations by removing $\omega$ dependence from the bulk Bloch Hamiltonian, and is a good approximation for small particles because $\omega$ varies faster in the polarizability than the Green's function. For larger particles the approximation becomes inaccurate at the light line $k_x = \pm \sqrt{\epsilon_B}\omega/c$.

\begin{figure*}[t!]
\centering
\includegraphics[]{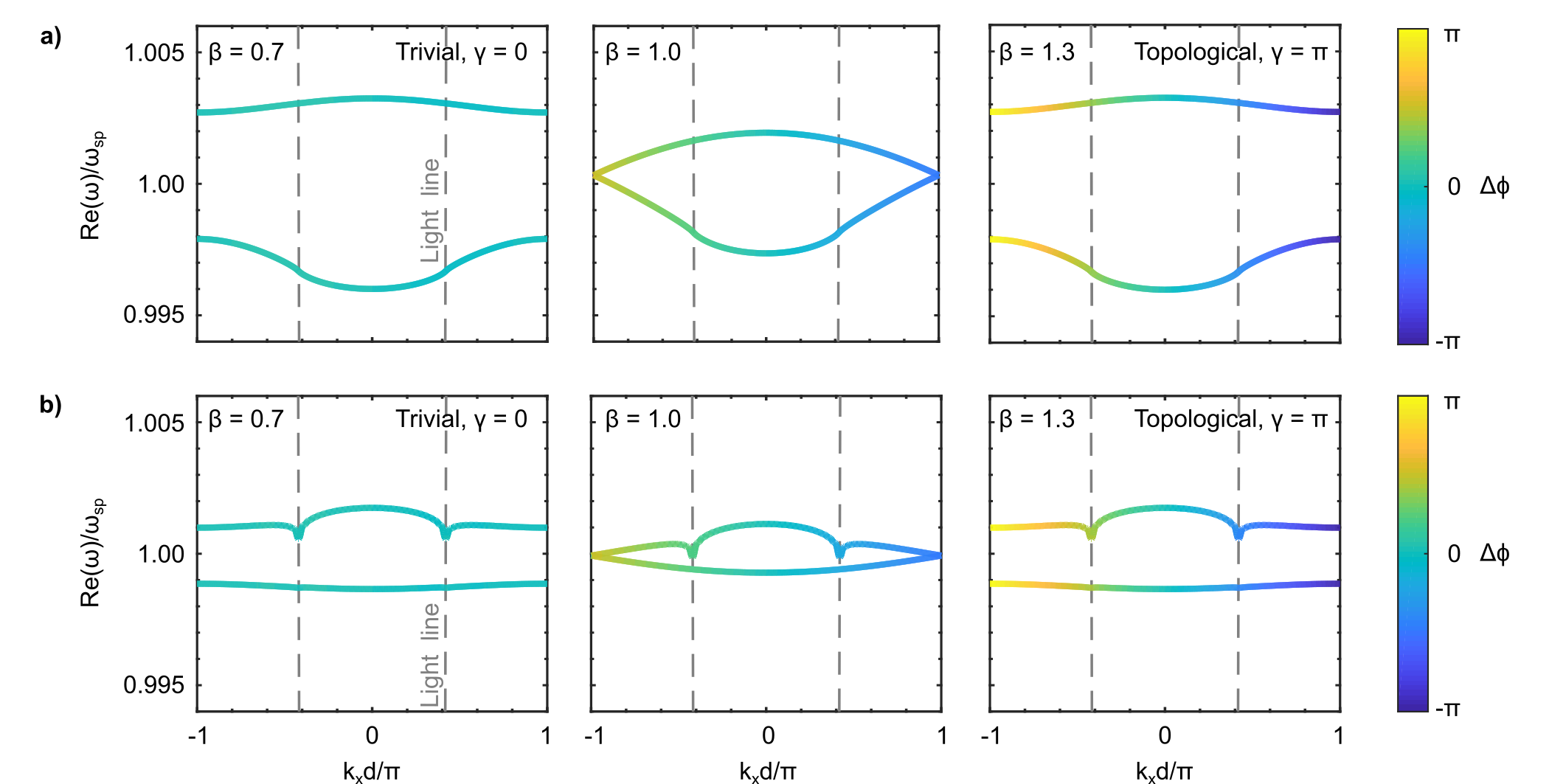}
\caption{Real part of the dispersion relation of the topological plasmonic chain with $a =5$\,nm and $k_{sp}d = 0.42\pi$ for the (a) longitudinal polarisation and (b) transverse polarisation, which corresponds to $d = 77$\,nm. Band colouring shows how $\phi$ changes for each band away from the centre of the BZ. Not depicted is that the upper and lower bands have a phase difference of $\pi$.}
\label{fig:Fig1}
\end{figure*}

For the one-dimensional chain the $x$, $y$ and $z$ components of equation~\ref{eqn:CDA} decouple. This leads to two distinct polarisations of the particles, longitudinal ($x$) and transverse ($y$ and $z$) in figure~\ref{fig:Fig0}(b). It is helpful to relabel particles by unit cell and sublattice, leading to a set of equations

\begin{align} \label{eqn:specificCDA}
\frac{d^3}{\alpha(\omega)}p^\nu_{A,n} &= \sum_{m\neq n}g^\nu(m-n)p^\nu_{A,m}  \nonumber \\ &+ \sum_{m}g^\nu(m-n+\beta/2)p^\nu_{B,m}, \nonumber \\
\frac{d^3}{\alpha(\omega)}p^\nu_{B,n} &= \sum_{m\neq n}g^\nu(m-n)p^\nu_{B,m} \nonumber \\ &+ \sum_{m}g^\nu(m-n-\beta/2)p^\nu_{A,m},
\end{align}

\noindent where $\nu = x,y,z$ represents the direction of polarisation and the linearised reduced Green's functions $g^\nu(r)$ are given by

\begin{align} 
g^x(r) &= 2 \frac{e^{ik_{sp}d|r|}}{|r|^3}\left[1 - ik_{sp}d|r| \right], \label{eqn:longrange}\\
g^{y,z}(r) &= - \frac{e^{ik_{sp}d|r|}}{|r|^3}\left[1 - ik_{sp}d|r| - (k_{sp}d)^2|r|^2 \right], \label{eqn:transverse}
\end{align}

\noindent with $r$ the spacing between particles divided by $d$ and $k_{sp} = \sqrt{\epsilon_B}\omega_{sp}/c$ the mangitude of the wavevector of light with the same frequency as the single particle SPR.  We refer to the third term in the transverse reduced Green's function (equation~\ref{eqn:transverse}) as `long range', as it decays proportional to the inverse of the particle separation. In addition we note the finite lifetime of the plasmons, which cause decoherence between the nanoparticles. Dipoles can only interact coherently if they have separation less than $c\tau/\sqrt{\epsilon_B}$, which we take into account by setting $g(r)$ to zero if $rd > c\tau/\sqrt{\epsilon_B}$.

This system is physically identical to one we examined in a previous work, but we use a slightly different model. By linearising the Green's functions we reduce the numerical difficulty of the model and allow ourselves to consider longer chains. The addition of the plasmon lifetime based cutoff removes physically unrealistic divergences at the light line.

In order to understand the topological properties of the system we study the bulk by considering an infinite chain. We relabel particles by unit cell and sublattice $A$ or $B$ and apply Bloch's theorem to arrive at the eigenvalue problem

\begin{equation}
\mathcal{G}^\nu(k_x)
\begin{pmatrix} p^\nu_A \\ p^\nu_B \end{pmatrix} = E(\omega,k_x)\begin{pmatrix} p^\nu_A \\ p^\nu_B \end{pmatrix},
\end{equation}

\noindent where $k_x$ is the $x$-component of the wavevector, the eigenvalue $E(\omega,k_x) = d^3/\alpha(\omega)$ and the components of the $2\times2$ Bloch Hamiltonian matrix are given by

\begin{align} \label{eqn:gvalues}
\mathcal{G}_{11}^\nu(k_x) = \mathcal{G}_{22}^\nu(k_x) &= \sum_{n\neq 0}g^\nu(n)e^{ik_xdn}, \nonumber \\
\mathcal{G}_{12}^\nu(k_x) &= \sum_{n}g^\nu(n+\beta/2)e^{ik_xdn}, \nonumber \\
\mathcal{G}_{21}^\nu(k_x) &= \sum_{n}g^\nu(n-\beta/2)e^{ik_xdn}.
\end{align} 

\noindent The matrix is non-Hermitian, which allows $E$ to take complex values. Solving the equation for $\omega$ gives the bulk dispersion relation of the chain. As an example we consider the case of $k_{sp}d/\pi = 0.42$, corresponding to $d=77$\,nm, in figure~\ref{fig:Fig1}, plotting the real part of the dispersion relation for the longitudinal (a) and transverse (b) polarisations with $\beta = 0.7$, $1$, and $1.3$. The band structure is symmetric in $k_x$ due to the inversion symmetry of the system.

The relevant topological number of this system is the Zak phase~\cite{Zak1989} using the periodic gauge~\cite{Resta2000,Ling2015}, given by

\begin{equation} \label{eqn:phi}
\gamma = \frac{\phi(\pi) - \phi(-\pi)}{2} \mod 2\pi,
\end{equation}

\noindent where $\phi(k_xd)$ is the relative phase between $p_A$ and $p_B$. Although a variation on the Zak phase has been proposed which takes into account the non-Hermiticity of the system~\cite{Wang2018b,Zhong2018b}, we will show that this Zak phase is sufficient here. In a system with bulk-edge correspondence a Zak phase of $\gamma = \pi$ ($\gamma = 0$) predicts the existence (non-existence) of topologically protected edge modes~\cite{Delplace2011}. The colouring of the bands in figure~\ref{fig:Fig1} shows how $\phi$ changes moving away from the centre of the Brillouin zone (BZ). We see that for $\beta <1$, $\gamma = 0$ and for $\beta >1$, $\gamma = \pi$. There is a bandgap closure at $\beta = 1$ for complex $\omega$ (imaginary part not shown in the figure), indicating a topological phase transition, and the colouring of the bands shows that the Zak phase is not an integer. Notably the closing of the gap in $\omega$ is equivalent to the closing of the gap in $E$ and it is therefore enough to consider the eigenvalues $E$ rather than $\omega$ when examining the topological properties of the system.

In this case the topological number is quantised by chiral symmetry, where a given Hamiltonian $\hat{H}$ satisfies the relation $\sigma_z\hat{H}\sigma_z = -\hat{H}$, where $\sigma_z$ is the Pauli spin matrix. We argued in a previous work~\cite{Pocock2018} that although $\mathcal{G}$ does not satisfy this relation, it is equal to a chirally symmetric matrix plus an identity term $\mathcal{G}_{11}I$. It therefore has the same eigenvectors as a chiral matrix and so the Zak phase, which is calculated from the eigenvectors, is equal to that of the chiral matrix $\mathcal{G}-\mathcal{G}_{11}I$. In chirally symmetric systems the eigenvalues come in positive and negative pairs, leading to a spectrum that is symmetric about $|E|=0$. The addition of the identity term shifts the bulk bands by $\mathcal{G}_{11}(k_x)$ in the complex plane.

However, a subtlety not noted was that although the bulk topological number is the same, bulk-edge correspondence is not necessarily preserved. The sublattice operators which guarantee the existence and protection of zero eigenvalue edge modes in the finite chiral system~\cite{Asboth2016} do not exist for the finite `trivially broken chiral' system.  Other studies have already argued that the existence of large enough $A$ to $A$ and $B$ to $B$ hopping causes the disappearance of edge modes in simple Hermitian and dipolar non-Hermitian systems~\cite{PerezGonzalez2018,Wang2018a}. In summary, chiral symmetry is the ingredient which protects the edge states, and although inversion symmetry or an added identity term in the bulk still quantises the Zak phase, they do not guarantee bulk-edge correspondence. We study the effect that `trivial chiral symmetry breaking' has on the plasmonic chain by considering finite systems which are forced to be chiral, setting the $A$ to $A$ and $B$ to $B$ hopping to zero artificially, and comparing these with the physical `full dipolar' model.

\begin{figure}[t!]
\centering
\includegraphics[]{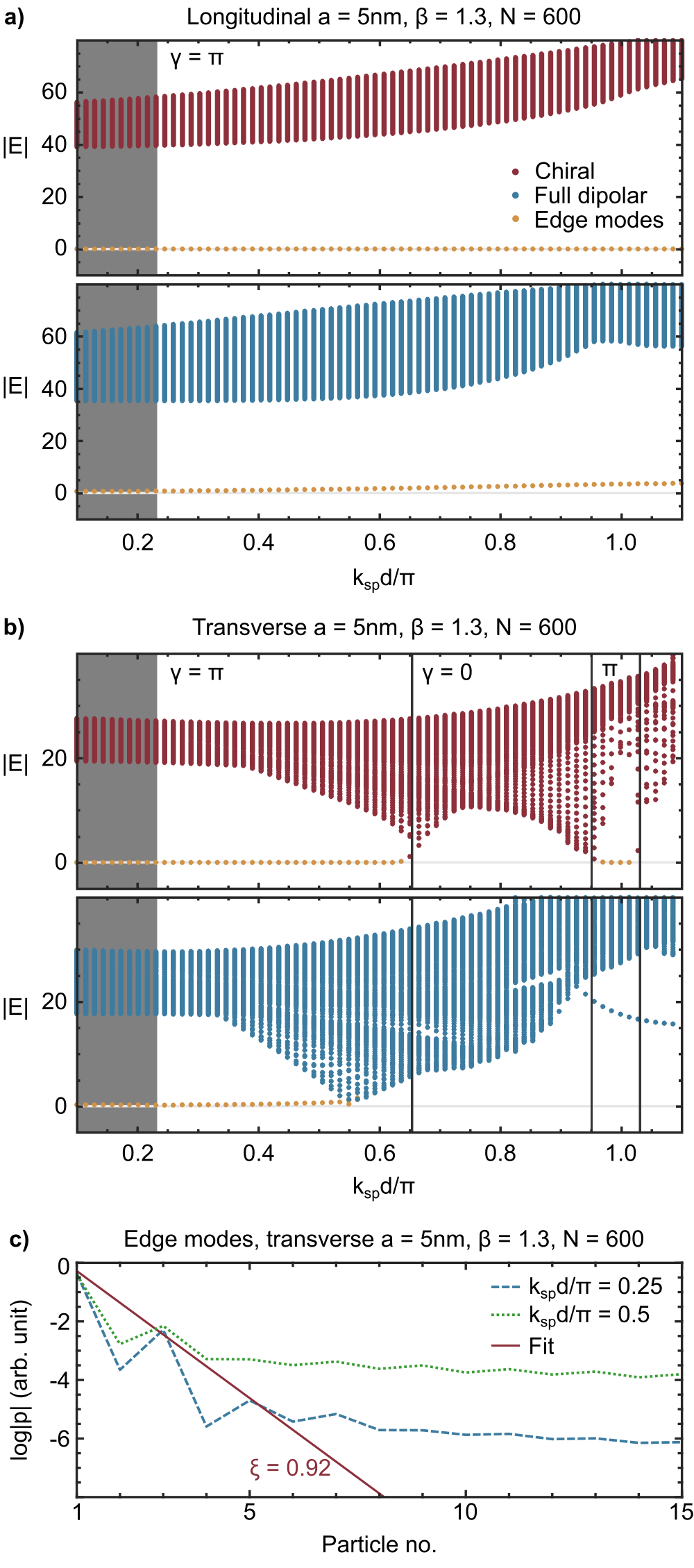}
\caption{Eigenvalues of the chiral (red) and full dipolar (blue) topological plasmonic chain with changing $k_{sp}d$ for the (a) longitudal and (b) transverse polarisations. Topologically protected edge modes are yellow. The dark grey area indicates the region where the CDA is not valid as the particles are too closely spaced. Vertical black lines indicate Zak phase transitions as predicted by the closing of the bulk gap. (c) Edge mode profiles of (b) blue for different choices of $k_{sp}d$.}
\label{fig:Finite}
\end{figure}

Figure~\ref{fig:Finite}(a) and (b) show $|E|$ for $N=600$ particle finite chains with $\beta = 1.3$ with changing $k_{sp}d$. Chiral bulk modes are shown in red and full dipolar modes in blue, with topologically protected edge modes coloured yellow. The dark grey region is where the spacing of the particles is too small for the CDA to be applicable, because at these shorter spacings higher order modes come into play. Plots of projected real and imaginary frequency $\omega$ for $\beta=1.4$ can be found in the SM. Fig~\ref{fig:Finite}(a) shows the longitudinal polarisation. For $\beta > 1$ the longitudinal bulk predicts a Zak phase of $\gamma = \pi$ and in this case we confirm that there are topologically protected edge modes fixed at $|E|=0$ in the chiral case, and in the full-dipolar case these modes slightly deviate from zero as the particle spacing increases. This is consistent with other works that suggest that the longitudinal full-dipolar case always features bulk-edge correspondence because $A$ to $A$ and $B$ to $B$ hoppings are `small enough' for all $k_{sp}d$.

Fig~\ref{fig:Finite}(b) shows the transverse polarisation, with vertical black lines marking the Zak phase transition according to bulk calculations.  In the chiral case we see that for small enough $k_{sp}d$, where the long range term in equation~\ref{eqn:longrange} is small, the Zak phase is equal to that of the longitudinal chain. As $k_{sp}d$ increases the Zak phase for $\beta$ above and below $1$ swaps at specific values of $k_{sp}d$ in what we previously called \textit{retardation induced topological phase transitions}, which are caused by the long range term. These phase transitions are non-Hermitian features which occur at exceptional points, where one of the off-diagonal terms in the Bloch Hamiltonian is zero. The exceptional points come in pairs symmetrically around $\beta = 1$; if $\mathcal{G}_{12}$ is zero for some value of $k_{sp}d$ for $\beta = 0.7$ then $\mathcal{G}_{21}$ is zero for the same $k_{sp}d$ for $\beta = 1.3$. Therefore for $\beta = 0.7$ the Zak phase is exactly opposite to $\beta = 1.3$ for all $k_{sp}d$. 

In the chiral case the existence and non-existence of zero edge modes are predicted perfectly by the Zak phase; as expected the chiral finite chain has bulk-edge correspondence. This proves that the Zak phase calculation is sufficient for this system in this parameter regime despite the non-Hermitian skin effect, which has lead some to propose a modification of the Zak phase in some $1D$ chiral systems~\cite{Zhong2018b,Wang2018b,Lee2018}.  In the full dipolar case, however, the bulk and edge modes intersect before the first phase transition and for higher $k_{sp}d$ there is no longer bulk-edge correspondence. In this work we will call this the \textit{bulk-edge correspondence breakdown}. We emphasise that the breakdown of bulk-edge correspondence in this system is a chiral symmetry breaking effect rather than a non-Hermitian effect, possibly because the system is not very non-Hermitian according to measurements of phase-rigidity in similar dipolar models~\cite{Wang2018a,Wang2018b}. BEC breakdown is not a topological phase transition: edge modes are lost in the $\beta>1$ case but not gained in the $\beta<1$ case (see supplementary material, SM), and the Zak phase does not change. The fact that this happens before the first Zak phase transition is consistent with the notion that the breakdown occurs because the long range term gets `too large' and therefore the $A$ to $A$ and $B$ to $B$ terms get too large, as the retardation induced phase transitions are also related to the long range term becoming important.

Another feature to note in the (blue) full dipolar case in figure~3(b) is the presence of modes outside the bulk in the $\beta = 1.3$ case for approximately $k_{sp}d/\pi > 0.95$. These modes also exist in other $\beta>1$ cases (see SM), and are localised to the edges of the chain. Importantly their existence or non existence does not line up exactly with the changing Zak phase, they have eigenvalues far from $|E|=0$, and they do not appear to be well protected from disorder. We therefore do not label these as topologically protected edge states and in this work still consider bulk-edge correspondence to be broken.

Fig~\ref{fig:Finite}(c) shows log plots of the dipole moments $|p|$ of two edge modes, one far from the bulk-edge correspondence breakdown at $k_{sp}d/\pi = 0.25$ and one close to the breakdown at $k_{sp}d/\pi = 0.5$. In a chiral system the edge modes are fully supported on one sublattice, which would in this case be the $A$ sublattice. Due to the chiral symmetry breaking edge modes spill into the $B$ sublattice, but we apply our exponential fit (red line) only to the $A$ sublattice. Unlike the usual description of SSH model phase transitions the edge modes are not fully exponential but rather appear to feature a highly localised edge part with localisation length $\xi$, which does not change as $k_{sp}d$ increases, and a bulk-like component that grows as $k_{sp}d$ increases. It appears that as the BEC breakdown approaches there is some mixing between the bulk and edge modes. This long bulk-like tail is related to the nature of the long range hopping. In figure~\ref{fig:Finite}(a) and (b) the edge modes are highlighted based on how close they are to zero, how localised the edge part is, and how much larger the edge part is compared to the average of the bulk part.

Although in the case of transverse polarisation chiral symmetry is strongly broken as $k_{sp}d$ becomes large enough, the system still has a quantised Zak phase due to its inversion symmetry represented by $\sigma_x \mathcal{G}(k_x) \sigma_x = \mathcal{G}(-k_x)$. Although this no longer corresponds to the presence of edge modes, there is some evidence that the topological properties of inversion symmetric systems manifest in other ways~\cite{Fu2007,Hughes2011}, and that there may be measureable consequences of topological phase independent of the existence of edge modes~\cite{Bello2018}.

Missing from our current description of BEC breakdown is an understanding of exactly how the $A$ to $A$ and $B$ to $B$ hoppings have to behave to break down the topological protection of the system. In order to better understand the effect that `trival' chiral symmetry breaking has on bulk-edge correspondence we take a slight mathematical diversion to study a much simpler non-Hermitian, next nearest neighbour extension to the SSH model. This will aid us when we return to the plasmonic chain afterwards.

\section{Non-Hermitian NNN SSH model}

\begin{figure}[t!]
\centering
\includegraphics[]{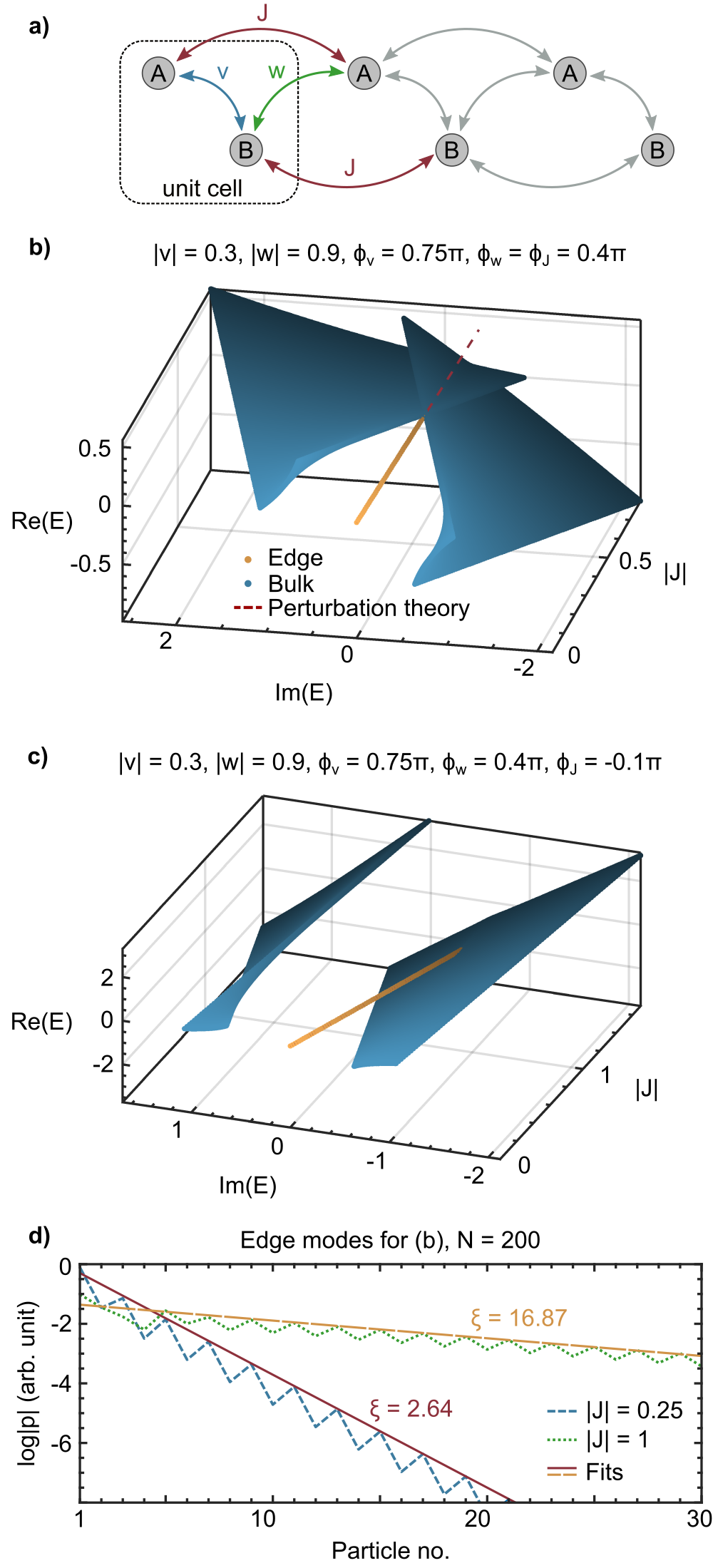}
\caption{(a) Diagram of next nearest neighbour SSH model with complex hopping $v$, $w$ and $J$. (b) and (c) Bulk (blue) and edge mode (yellow) eigenvalues of the non-Hermitian NNN SSH model for changing values of $|J|$, for different choices of hopping parameters and phases in (b) and (c). Modes with larger $|J|$ values are darker. The red dashed line is given by perturbation theory (see SM). (d) Edge mode profiles for the choice of parameters given in (c) for different choices of $|J|$. For real and imaginary projections see SM.}
\label{fig:NNN}
\end{figure}

We consider a next nearest neighbour (NNN) SSH model much like in a work by P\'{e}rez-Gonz\'{a}lez \textit{et al.}~\cite{PerezGonzalez2018} with the addition of symmetric complex valued hopping as in figure~\ref{fig:NNN}(a), which makes the model non-Hermitian. This is a simplified version of a model studied by Li \textit{et al.}~\cite{Li2019} Intracell hopping is given by $v=|v|e^{i\phi_v}$, intercell hopping by $w=|w|e^{i\phi_w}$ and $A$ to $A$ and $B$ to $B$ hopping by $J=|J|e^{i\phi_J}$. The Hamiltonian is given by

\begin{align}
\mathcal{H}= \sum_{n=1}^{N} &v\left[|n,A\rangle\langle n,B| + H.c. \right] \nonumber \\
+\sum_{n=1}^{N-1} &w\left[|n+1,A\rangle\langle n,B| + H.c. \right] \nonumber \\
+\sum_{n=1}^{N-1} &J[|n+1,A\rangle\langle n,A|+|n+1,B\rangle\langle n,B| + H.c.],
\end{align}

\noindent where $H.c.$ is the Hermitian conjugate. This model has bulk Bloch Hamiltonian

\begin{equation}
\mathcal{H}_{bulk}(k)=\begin{pmatrix}
2J\cos(k) & v+e^{-ik}w \\
v+e^{ik}w & 2J\cos(k) \end{pmatrix}.
\end{equation}

\noindent This Hamiltonian  has similarities to $\mathcal{G}$ in that it has inversion symmetry and chiral symmetry broken by an identity term, in this case proportional to the complex variable $J$. In fact, the chiral system that this model shares bulk eigenvectors with is exactly the non-Hermitian SSH model studied by Lieu~\cite{Lieu2018a} and the system has Zak phase $\gamma=\pi$ ($\gamma=0$) when $|w|>|v|$ ($|w|<|v|$), with topological phase transition at $|w| = |v|$.

Solving the eigenvalue problem gives the bulk eigenvalues,

\begin{equation}
E_{bulk} = 2J\cos(k) \pm \sqrt{v^2+w^2+2vw\cos(k)},
\end{equation}

\noindent while the variation of edge eigenvalues from the chiral system's $E_{edge}=0$ caused by the addition of small $J$ is given by perturbation theory (see SM), which confirms the fit produced by P\'{e}rez-Gonz\'{a}lez \textit{et al.}~\cite{PerezGonzalez2018} in the Hermitian case,

\begin{equation} \label{eqn:PT}
E_{edge} = -2J\frac{v}{w}.
\end{equation}

\begin{figure}[t]
\centering
\includegraphics[]{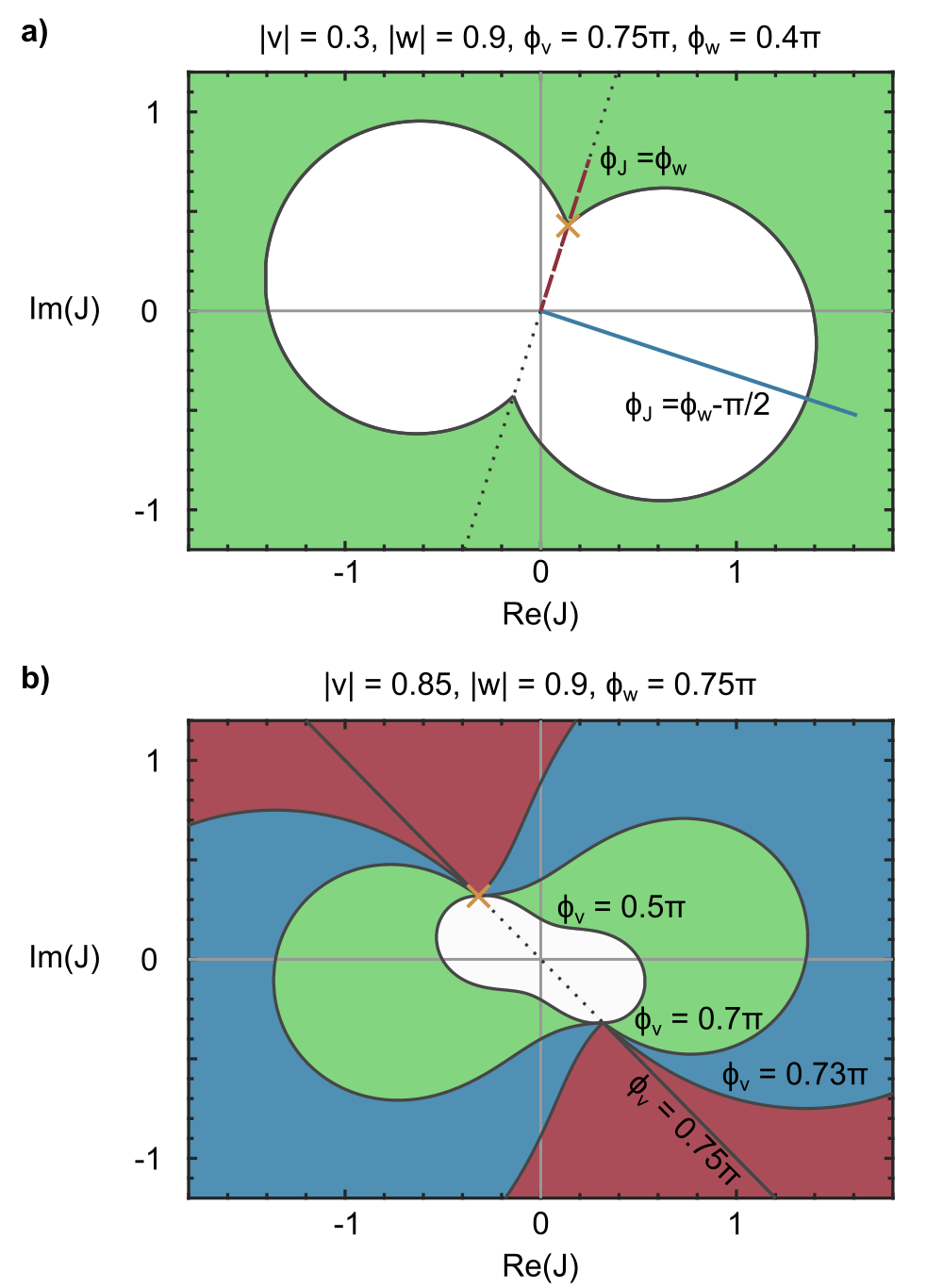}
\caption{(a) BEC phase diagram with predicted region of bulk-edge correspondence (white) and its breakdown (green) for parameters of the chain from fig~\ref{fig:NNN}, with $\phi_J$ from fig~\ref{fig:NNN}(a) (red dashed line) and fig~\ref{fig:NNN}(b) (blue line), yellow cross at $w/2$ and dotted line along $\phi_w$. (b) BEC phase diagrams for $|v| \sim |w|$ and varying $\phi_v$. The red, blue and green colourings represent different regions where BEC breakdown occurs for different choices of $\phi_v$, with red for $\phi_v=0.73\pi$, red and blue for $\phi_v = 0.7\pi$, and finally all of red, blue and green for $\phi_v = 0.5\pi$}
\label{fig:phasediagrams}
\end{figure}  

Figures~\ref{fig:NNN}(b) and (c) show the evolution of finite chain eigenvalues $E$ as $|J|$ increases for chains with $N=200$ particles with $|v|=0.3$, $|w|=0.9$, $\phi_v=0.75\pi$, $\phi_w=0.4\pi$ and two different choices of phase for $J$, $\phi_J=0.4\pi$ in (b) and $\phi_J=-0.1\pi$ in (c). Bulk modes are coloured blue and edge modes coloured yellow, with a red dashed line showing the predicted path of edge modes according to equation~\ref{eqn:PT}. Real and imaginary projected plots are provided in the supplementary material. As expected we see that edge modes enter the bulk for some value of $|J|$ and then disappear, marking the breakdown of bulk-edge correspondence in the system. Figure~\ref{fig:NNN}(b) shows a choice of parameters where the bulk bands move to close and cross the path of the edge modes, while figure~\ref{fig:NNN}(c) shows the case where the bulk does not close, but the edge modes move into one of the bulk bands and thereby destroy bulk-edge correspondence. From these figures we see that there are the two phenomena that appear to govern the destruction of bulk-edge correspondence for long chains: the movement of the bulk as in (b), and the movement of the edge modes as in (c).  In fact, there is a third case where the bands close due to finite size effects, but we ignore this for now as we cannot quantify it with the bulk Hamiltonian and its effect can be reduced by increasing the number of particles in the chain. For any set of parameters $v$, $w$ and $J$, BEC breakdown is caused by a combination of these effects.

Figure~\ref{fig:NNN}(d) shows edge modes of figure~\ref{fig:NNN}(c) for particular choices of $|J|$. As before they delocalise as we approach the BEC breakdown, but unlike the plasmonic topological insulator case they are completely exponential and as $|J|$ increases so does the localisation length $\xi$, until they are no longer localised.

Ignoring finite size effects we can find where the bulk and edge modes first intersect by solving $E_{edge}(k) = E_{bulk}(k)$ for $J(k)$. We then plot this value of $J$ for each $k$ in the BZ, constructing the phase diagrams as in figure~\ref{fig:phasediagrams}. Figure~\ref{fig:phasediagrams}(a) shows the BEC phase diagram for the parameters used in figure~\ref{fig:NNN}(b) and (c). If the system starts at the origin with $|J|=0$ and moves in a straight line away as the chirality breaking parameter $|J|$ increases, the edge modes and bulk meet when we cross the black line from the white region to the green region. Therefore the white region corresponds to the values of $J$ with bulk-edge correspondence and the green region corresponds to values where it is broken. The red dashed and blue solid lines correspond to choices of $\phi_J$ from figures~\ref{fig:NNN}(b) and (c) respectively. We can see that for figure~\ref{fig:NNN}(b) the bulk bands are shifted directly towards $|E|=0$, whereas in figure~\ref{fig:NNN}(c) they do not move towards $|E|=0$, so that the slower movement of the edge modes causes BEC breakdown. This illustrates that breakdown occurs for different degrees of chiral symmetry breaking, $|J|$, depending on the direction the bands are shifted due to choice of parameter $\phi_J$. For $|v| << |w|$, as in figure~\ref{fig:phasediagrams}(a), the phase corresponding to the smallest $|J|$ at which BEC breakdown occurs is $\phi_J = \phi_w$ and the transition happens at $J = w/2$ (yellow cross). Comparing figures~\ref{fig:NNN} and~\ref{fig:phasediagrams}(a) we see that when $\phi_J \sim \phi_w$ BEC breakdown is dominated by bulk movement and when $\phi_J$ is far from $\phi_w$ it is dominated by edge movement. In the Hermitian model $\phi_J = \phi_w$ always. Non-Hermiticity allows for complex $J$ and therefore leads to different behaviours depending on the `path' of the bulk in complex space. 

Figure~\ref{fig:phasediagrams}(b) shows the case where $|v|$ is similar but still smaller than $|w|$, for different values of $\phi_v$. The colouring represents different regions where breakdown occurs for different choices of $\phi_v$, with red for $\phi_v=0.73\pi$, red and blue for $\phi_v = 0.7\pi$, and finally all of red, blue and green for $\phi_v = 0.5\pi$. As $\phi_v$ gets closer to $\phi_w$ the region of bulk-edge correspondence becomes larger, until at $\phi_v = \phi_w$ it theoretically only breaks down when $\phi_J = \phi_w$. However, for $\phi_J$ further from $\phi_w$ the breakdown is dominated by finite size effects which shrink the BEC region. In this case when $\phi_v$ is not similar to $\phi_w$ we also see that the shortest path in $J$ space is not necessarily along $\phi_J$.

We briefly summarise the results which will be useful for our study of the plasmonic chain. Increasing chiral symmetry breaking causes both the bulk and edge modes to be shifted in complex space until one collides with the other, at which point BEC breakdown occurs. We have observed a system where depending on the choices of parameter either the bulk or edge mode movement can be made to act as the dominant effect. The parameter in question chooses the direction in which the bands are shifted in complex space, which has a significant effect on exactly how much chiral symmetry breaking causes BEC breakdown. This is an important question only in a non-Hermitian system where eigenvalues are complex, because in a Hermitian system the `direction parameter' is fixed along the real line.

\section{BEC breakdown in the plasmonic chain}

Previous works have discussed the fact that a breakdown in bulk-edge correspondence occurs in the transverse case when the long range term in the Green's function becomes large enough that $A$ to $A$ and $B$ to $B$ hoppings are comparable to $A$ to $B$ and $B$ to $A$ hoppings~\cite{Wang2018a,Wang2018b}. In the following we elaborate on what these requirements actually are and how we can find where this occurs by considering bulk terms.

As discussed in the previous section, depending on the parameters of a system BEC breakdown can be dominated by the movement of the bulk or of the edge modes. Although it is difficult to exactly align parameters of the plasmonic system with parameters from the NNN system, we can see from plots like figure~\ref{fig:Finite}(b) and others (see SM) that the edge modes do not move far from zero before they enter the bulk. We note that the edge modes do move a little more when $\beta$ is further from $\beta = 1$, but nevertheless conclude that the movement of the bulk dominates and that it is reasonable to approximate $E_{edge} = 0$. The bulk bands $E_{bulk}(k_x)$ are given by the off-diagonal terms of the bulk Bloch Hamiltonian $\pm\sqrt{\mathcal{G}_{12}(k_x)\mathcal{G}_{21}(k_x)}$ plus a shift in complex space by the chiral symmetry breaking on-diagonal term $\mathcal{G}_{11}(k_x)$. In the Hermitian case~\cite{PerezGonzalez2018} one of the bands can cross zero when, for some value of $k_x$ in the Brillouin zone, the magnitude of the chirality breaking shift is equal to that of the chiral bands,

\begin{equation} \label{eqn:HermitianZero}
|\mathcal{G}_{11}(k_x)| = |\sqrt{\mathcal{G}_{12}(k_x)\mathcal{G}_{21}(k_x)}|.
\end{equation} 

\noindent As we saw in the previous section, when considering a non-Hermitian system we must be careful because depending on the parameters the bands are not necessarily shifted towards $|E|=0$. The plasmonic system's equivalent parameter to the NNN system's $\phi_J$ is the phase of $\mathcal{G}_{11}$, which differs in that it has $k_{sp}d$ dependence. In much the same way that $k_x$ was the parameter for the breakdown value of $J$ in figure~\ref{fig:phasediagrams}, we are interested in where in the BZ, given by $k_x$, the bulk bands of the plasmonic system first cross $|E|=0$, in order to identify which direction the phase of $\mathcal{G}_{11}$ points. By observing bulk band structures we see that dips at the light line lead to the zero-line crossing as $k_{sp}d$ increases, because these dips give the minimum $|E|$ for one of the bands. We also observe that at the light line the phases of $\mathcal{G}_{11}(k_x)$ and one of $\pm\sqrt{\mathcal{G}_{12}(k_x)\mathcal{G}_{21}(k_x)}$ are approximately equal, possibly related to the phase contribution given by $\exp(ik_{sp}d\pm ik_xd) = 1$ at the light lines.

These observations tell us that the plasmonic system is in a situation roughly equivalent to the NNN $\phi_J \approx \phi_w \approx \phi_v$ case. The edge mode eigenvalues don't move significantly, and for certain $k_x$ the bulk bands move almost directly towards $|E|=0$ as $k_{sp}d$ increases. This means that for certain $k_x$ when the shift $\mathcal{G}_{11}(k_x)$ is equal in magnitude to the chiral bands $\sqrt{\mathcal{G}_{12}(k_x)\mathcal{G}_{21}(k_x)}$ we expect the bands to cross $|E|=0$. In light of this we define a measure of non-chirality for this system,

\begin{equation}
\eta = \max\displaylimits_{k_x\in\mathrm{BZ}}\frac{|\mathcal{G}_{11}(k_x)|}{|\sqrt{\mathcal{G}_{12}(k_x)\mathcal{G}_{21}(k_x)}|}.
\end{equation}

\begin{figure}[t]
\centering
\includegraphics[]{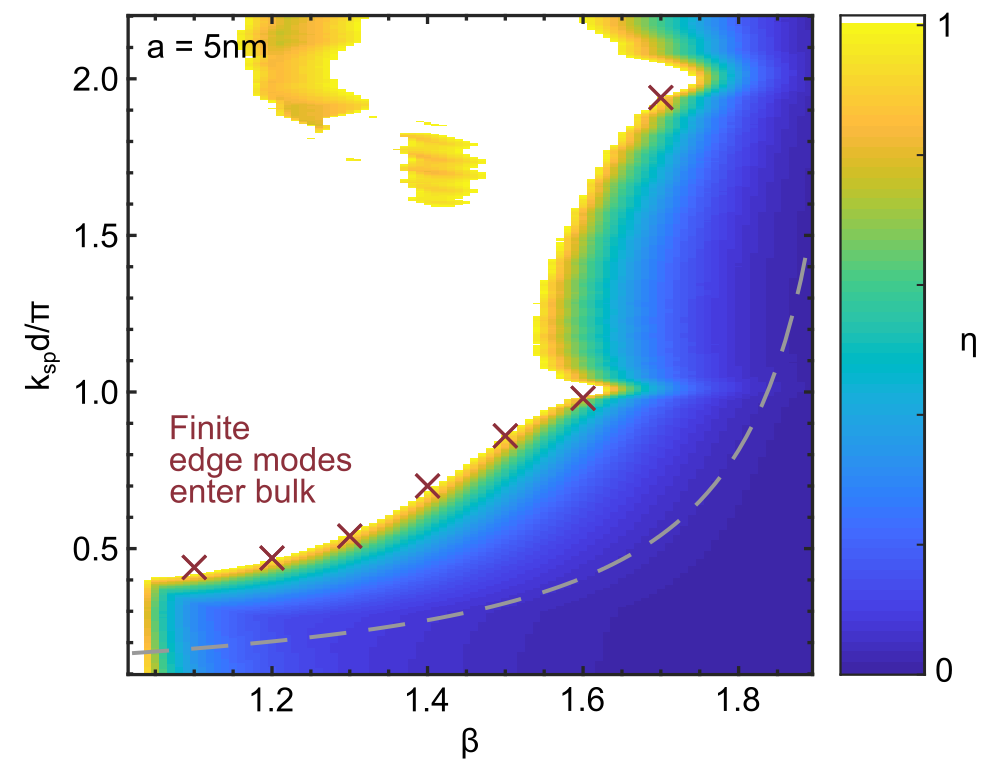}
\caption{$\eta$ for the transverse modes of the chain, for different $\beta$ and $k_{sp}d$, which saturates so that values of $\eta>1$ are white. Red crosses correspond to the locations where edge modes enter the bulk in finite 600 particle chains, possibly indicating the BEC breakdown. The grey dashed line corresponds to the minimum value of $k_{sp}d/\pi$ for which the CDA is a good approximation.}
\label{fig:eta}
\end{figure}

\noindent We plot $\eta$ in figure~\ref{fig:eta} for changing $k_{sp}d$ versus $\beta$ for the transverse modes of the chain, overlaid with red crosses where the edge modes of finite chains of $N=600$ enter the bulk. The colour saturates so that white corresponds to values of $\eta>1$. The grey dashed line is the minimum value of $k_{sp}d$ for which the CDA is a good approximation. From the definition, when $\eta = 0$ the system is fully chiral. We consider $\eta$ by once again fixing $\beta$, starting with small $k_{sp}d$, then increasing to see how the BEC breakdown occurs as we do so. For small $k_{sp}d$, when $\eta <1$ the bulk cannot have crossed the zero line, so we expect the edge modes to still exist and to have bulk edge correspondence. Approximately when $\eta = 1$ one of the bulk bands touches the zero line and therefore the edge modes have entered the bulk, with some small disagreement due to the $E_{edge} = 0$ and phases being equal at the light lines approximations. Finally when $\eta > 1$ we expect BEC to have broken down because the edge modes have already entered the bulk. In figure~\ref{fig:eta} we see that $\eta=1$ at roughly the same $k_{sp}d$ for which BEC breakdown occurs in the finite systems, again shown as red crosses. This supports the notion that $\eta$ is a good measurement of non-chirality for the system.

In figure~\ref{fig:eta} we see that after the lowest values of $k_{sp}d$ such that $\eta > 1$, increasing $k_{sp}d$ sometimes leads to a region where $\eta <1$, such as the yellow islands surrounded by white. We do not necessarily expect the return of exact bulk-edge correspondence, confirmed by finite chain simulations like those in figure~\ref{fig:Finite}(b). However, if $\eta$ ever returns to $\eta = 0$ then BEC must return because the system must be perfectly chiral. This means it is difficult to claim anything certain about these particular cases. Peaks at $k_{sp}d/\pi = n$ where $n$ is an integer are due to the light lines meeting at the edge of the BZ and combining to cross the $|E|=0$ line for lower $k_{sp}d$. Finally, for the longitudinal case in this region and all regions examined we find $\eta<1$, which agrees with the fact that we always expect bulk-edge correspondence with this polarisation.

From both the bulk measure $\eta$ and finite systems (red crosses) we observe that for a given value of particle radius, here $a=5$\,nm, as the period of the chain increases, BEC breaks down in agreement with our discussion of the breaking of chiral symmetry. The measure $\eta$ is a useful measure of the non-chirality of the system which can be used as a map for experimentalists to search for parameters of $k_{sp}d$ that can be expected to exhibit bulk-edge correspondence. In the figure $k_{sp}d/\pi=1$ corresponds to $d=184$\,nm, so for example a chain with $\beta=1.4$ can be expected to retain bulk-edge correspondence up to approximately $d=120$\,nm. For larger values of $\beta$, the value of the period for which BEC breakdown occurs increases. Hence, it would be experimentally favourable to choose larger values of $\beta$ and larger periods $d$ to observe topological edge modes in these plasmonic chains as the particles would not need to be very closely spaced, relaxing fabrication constraints. On the other hand, regarding particle size, we expect qualitatively similar results for particles up to $20$\,nm radius. Here radiative effects would be more prominent but the topological properties would be qualitatively the same as we describe in this paper for smaller particles~\cite{Pocock2018}. Experimental techniques such as cathodoluminescence spectroscopy and non-linear light generation can be used to probe topological band structures in photonics~\cite{Kruk2018,Peng2019}. Plasmonic system with similar SSH-like physics have also been studied experimentally~\cite{Poddubny2014,Sinev2015}. Other dipolar systems such as chains of cold atoms or phonons in SiC would also be expected to exhibit similar topological behaviour because they exhibit the same type of hoppings~\cite{Wang2018a,Wang2018b}.

Beyond the topological plasmonic chain, individual systems featuring chiral symmetry breaking must be examined on a case-by-case basis to understand where BEC breakdown occurs, by considering the behaviour of the edge modes and bulk. As demonstrated by the above study one can apply some of the knowledge gained from the simple non-Hermitian NNN SSH model to develop a measure for a more complicated system.

\section{Conclusion}

In this article we have broadened the discussion of bulk-edge correspondence in non-Hermitian systems, elaborating on the question of what happens when chiral symmetry is broken in one dimensional systems. We have shown that the question of how strictly chiral symmetry must be obeyed in order to observe topological protection is an important one if we wish to be careful about real world topological insulators like the topological plasmonic chain.

We recalled the model for a chain of metallic nanoparticles with alternating spacing and discussed how such a chain has been shown to exhibit topological properties such as a quantised Zak phase and topological protection of edge modes. Photonic systems like the plasmonic chain are natural non-Hermitian systems and can provide a valuable tool for theoretical and experimental exploration of topological insulators. The chain was shown to exhibit a breakdown of this bulk-edge correspondence in the transverse polarised case for large spacing, although the existence of predicted `retardation induced phase transitions' could still be measured as a kind of weak topological insulator with inversion symmetry.

We examined the non-Hermitian, next nearest neighbour SSH model to provide some basic intuition for the phenomenon of BEC breakdown, which is shown to be caused by the movement of the bulk, the edge modes and finite size effects, the first two of which could be visualised with phase diagrams. This informed our study of the plasmonic chain, where we defined a measure of chiral symmetry breaking $\eta$ to find a parameter regime for experimentalists to search for topologically protected transverse edge modes in the system.

Beyond the models considered in this article we have provided a framework for assessing the breakdown of bulk-edge correspondence in systems where the movement of the bulk movement dominates and is directed towards zero in eigenspace, or the bulk and edge modes move predictably.

\section{Acknowledgments}

SRP thanks S. Lieu for helpful discussions on topological insulators, and acknowledges funding from EPSRC. PAH acknowledges the Gordon and Betty Moore Foundation. VG acknowledges the Spanish Ministerio de Economia y Competitividad for financial support through the grant NANOTOPO (FIS2017-91413-EXP), and also Consejo Superior de Investigaciones Cientficas (INTRAMURALES 201750I039).

\newpage

\section{Supplementary Material}

\subsection*{More bulk boundary breakdown plots}

\subsubsection*{Behaviour of $\beta < 1$ systems}

\begin{figure}[t!]
\centering
\includegraphics[]{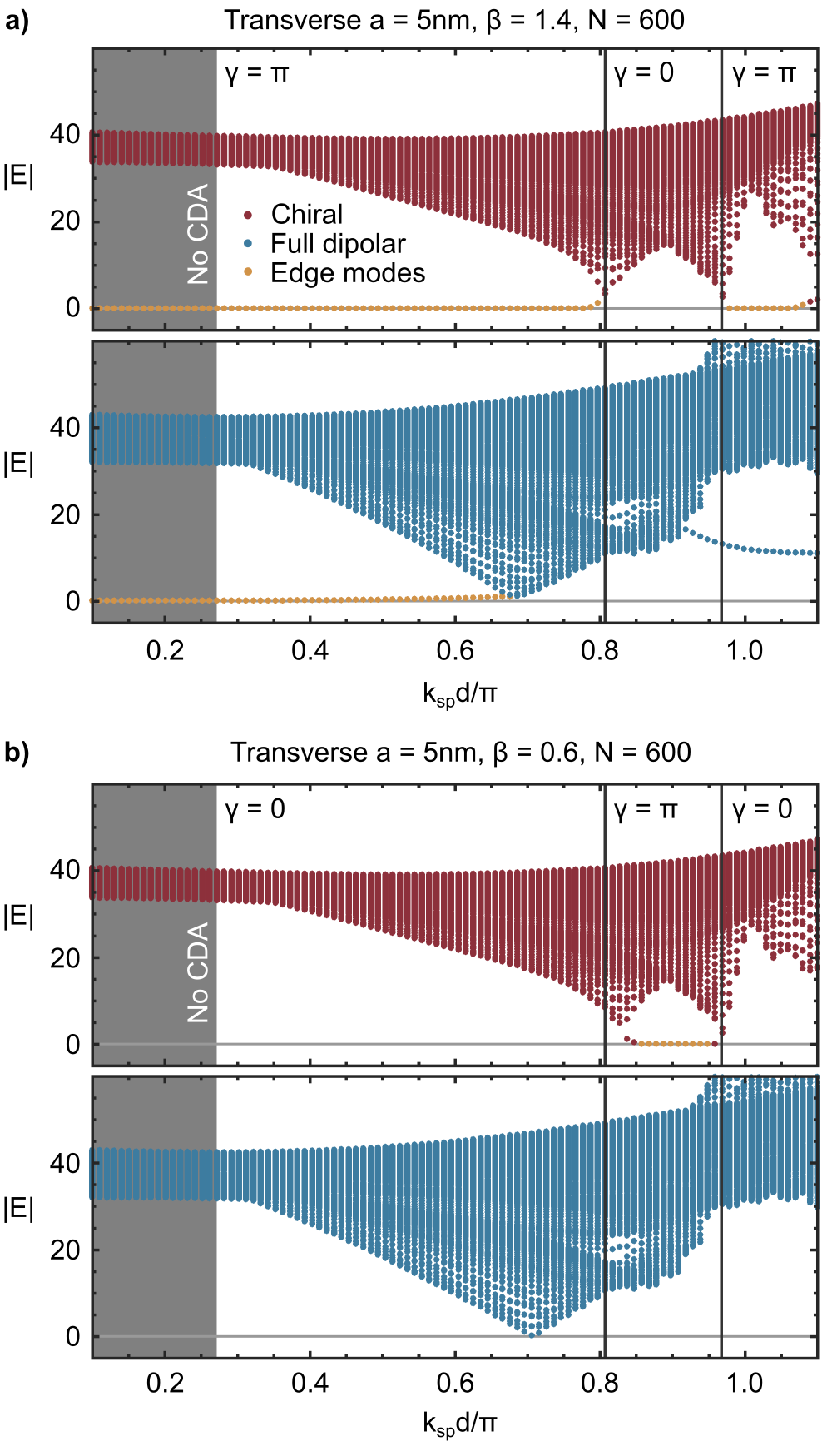}
\caption{Eigenvalues of the chiral (red) and full dipolar (blue) topological plasmonic chain with changing $k_{sp}d$ for the transverse polarisation with (a) $\beta = 1.4$ and (b) $beta = 0.6$. Topologically protected edge modes are yellow. The dark grey area indicates the region where the CDA is not valid as the particles are too closely spaced. Vertical black lines indicate Zak phase transitions as predicted by the closing of the bulk gap.}
\label{fig:betalessone}
\end{figure}

Figure~\ref{fig:betalessone} shows $|E|$ for $N=600$ particle finite plasmonic chains with (a) $\beta = 1.4$ and (b) $\beta = 0.6$ with changing $k_{sp}d$. Zak phase transitions occur for the chain at $\beta = 1$ and at values symmetric around $\beta = 1$, so that symmetrically either side of $\beta=1$ the Zak phase is exactly opposite. Comparing the chiral (red) case for (a) and (b) we see that topologically protected edge modes exist in the appropriate regions given the changing Zak phase. The non-chiral (blue) full dipolar case highlights that although BBC breakdown leads to the disappearance of edge modes in the $\beta>1$ case (a), it does not lead to the appearance of edge modes for the opposite $\beta<1$ in (b). In addition to this we see that for $\beta = 1.4$ the movement of the bulk dominates BEC breakdown in (a), where the edge modes are nearly zero until they enter the bulk for the first time.

As mentioned in the main text, we observe in the (blue) full dipolar case the presence of modes outside the bulk in the $\beta = 1.4$ case (which do not exist for $\beta = 0.6$) for approximately $k_{sp}d/\pi > 0.9$. As discussed in the main text these modes and are localised to the edges of the chain but their existence does not agree with the Zak phase, they are far from $|E|=0$, and do not appear to be well protected from disorder. As such we do not label these as topologically protected edge states and consider the system to still have broken bulk-edge correspondence. Although it appears that these are approaching the $|E|=0$ point in this plot these modes actually move back up into the bulk as $k_{sp}d/\pi$ increases. This return and loss of modes outside of the band which do not correspond to the Zak phase continues to happen as $k_{sp}d/\pi$ increases.

\subsubsection*{Real and imaginary parts}

\begin{figure}[t!]
\centering
\includegraphics[]{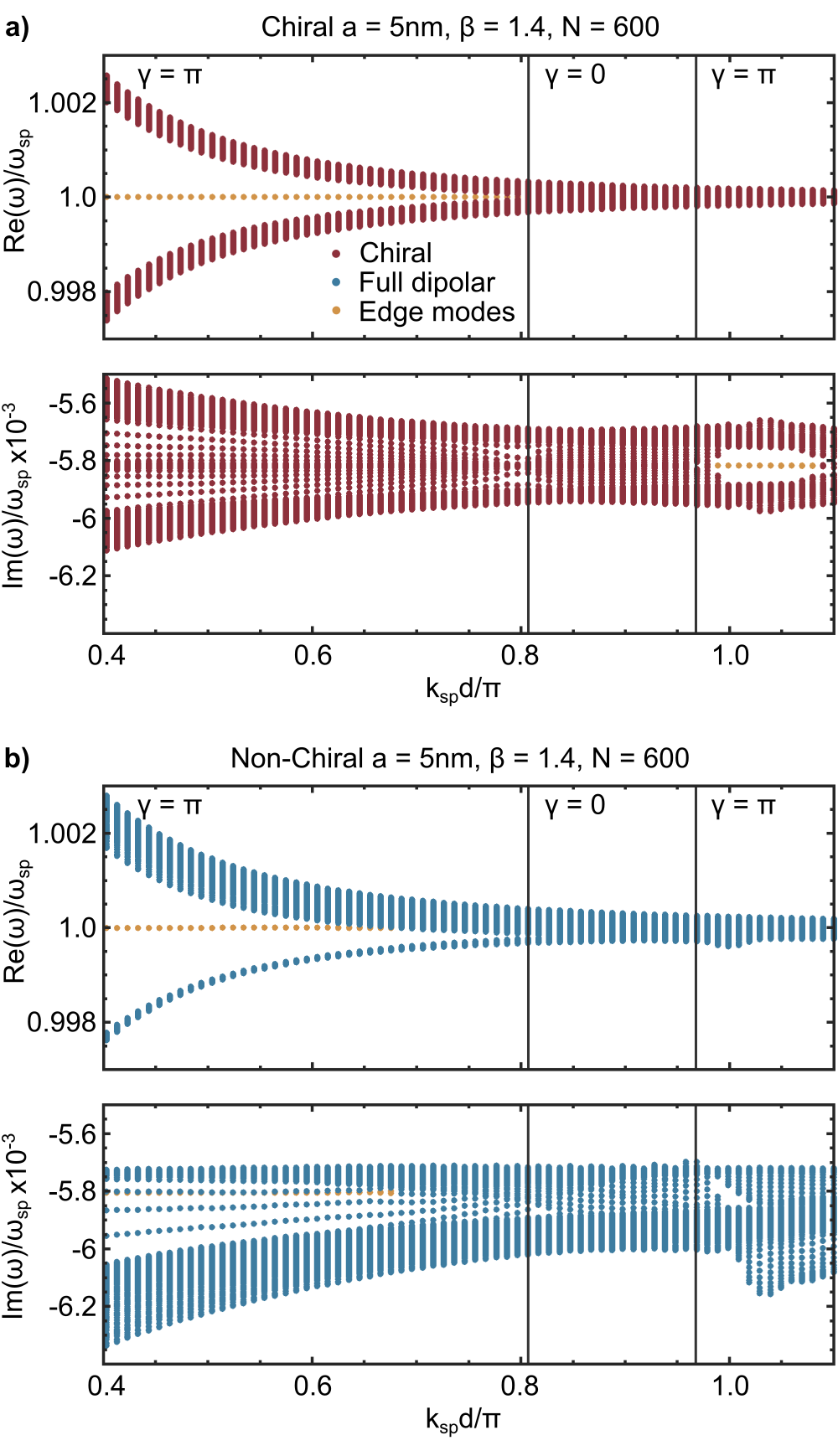}
\caption{Frequency values of the chiral (red) and full dipolar (blue) topological plasmonic chain with changing $k_{sp}d$ for the transverse polarisations. Topologically protected edge modes are yellow. Vertical black lines indicate Zak phase transitions as predicted by the closing of the bulk gap.}
\label{fig:reimparts}
\end{figure}

Figure~\ref{fig:reimparts} shows the real and imaginary parts of the frequency $\omega$ for figure~\ref{fig:betalessone}(a), similar to 3(b) in the main text but for $\beta=1.4$, using $E = d^3/\alpha(\omega)$. In the non-chiral (blue) case edge modes disappear as the bulk crosses their path, as shown in the $|E|$ plots featured in the main text.

\begin{figure}[t!]
\centering
\includegraphics[]{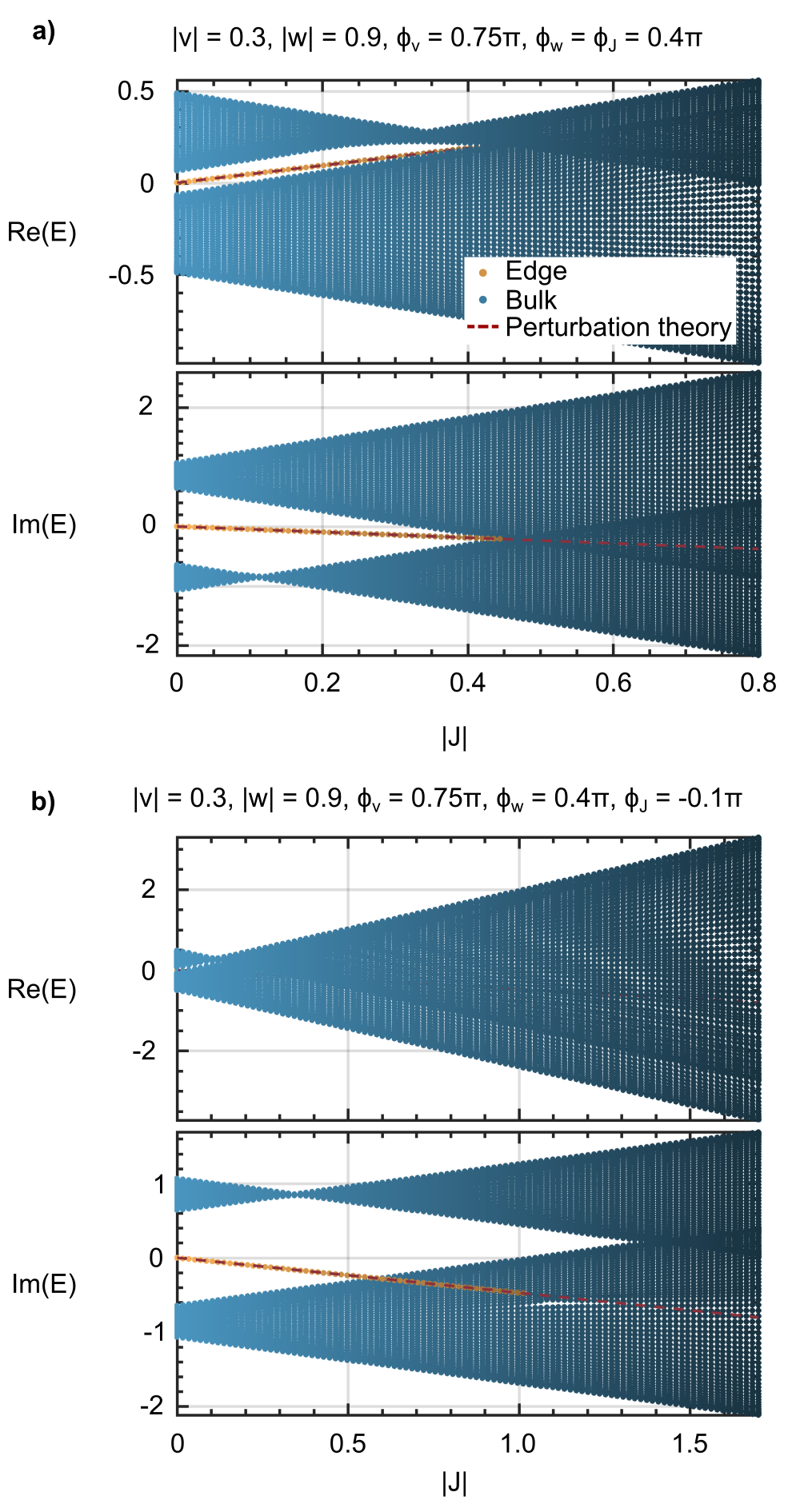}
\caption{(a) and (b) Bulk (blue) and edge mode (yellow) eigenvalues of the non-Hermitian NNN SSH model for changing values of $|J|$, for different choices of hopping parameters and phases corresponding to figures 4(b) and (c) respectively in the main text.}
\label{fig:reim}
\end{figure}

Figure~\ref{fig:reim} shows the real and imaginary parts of the eigenvalues for figure 4(b) and (c) in the main text. For the real part in Figure~\ref{fig:reim}(b) the bulk obscures the edge modes, but BEC correspondence doesn't break down until the edge modes disappear in the imaginary part plot.

\subsection*{Perturbation theory for NNN edge state energies}

We start with a finite SSH model with complex hoppings $v$ and $w$ with symmetric but non-Hermitian Hamiltonian of the form,

\begin{align}
\mathcal{H}^0 = &v\sum_{n=1}^{N} \left[|n,A\rangle\langle n,B| + H.c. \right] \nonumber \\
+ &w\sum_{n=1}^{N-1} \left[|n+1,A\rangle\langle n,B| + H.c. \right].
\end{align}

\noindent We treat the addition of next nearest neighbour hoppings as a perturbation by an operator $\mathcal{J}$, which features $A$ to $A$ and $B$ to $B$ hoppings and is of the form,

\begin{equation}
\mathcal{J} = J\sum_{n=1}^{N-1}[|n+1,A\rangle\langle n,A|+|n+1,B\rangle\langle n,B| + H.c.].
\end{equation}

We wish to find what happens to the zero energy edge modes of $\mathcal{H}^0$ as we increase the strength of the next nearest neighbour hopping, by examining the matrix $\mathcal{H}^1 = \mathcal{H}^0 + \mathcal{J}$, where we treat $J$ as the small parameter which turns on the next nearest neighbour hopping. Since $\mathcal{H}^0$ is symmetric its right and left eigenvectors are related simply,

\begin{align}
\mathcal{H}^0 \mathbf{v}^0_n &= E^0_n \mathbf{v}^0_n, \\
(\mathbf{v}^0_n)^T \mathcal{H}^0  &= E^0_n (\mathbf{v}^0_n)^T,
\end{align}

\noindent with the orthogonality condition that $(\mathbf{v}^0_m)^T \mathbf{v}^0_n = 0$ if $ m\neq n$. We are interested in looking at the edge modes which have $E^0 = 0$, so we can work through the usual perturbation theory using the symmetry properties to show that, to first order, the energy of the perturbed edge modes will be given by

\begin{equation} \label{eqn:pert}
E^1 = \frac{(\mathbf{v}^0)^T\mathcal{J} \mathbf{v}^0}{(\mathbf{v}^0)^T\mathbf{v}^0 }.
\end{equation}

According to section 1.5.6 of \textit{A Short Course On Topological Insulators}~\cite{Asboth2016}, in the thermodynamic limit the left and right edge modes of the unperturbed SSH chain are approximately given by the following:

\begin{equation}
|L\rangle = \sum\displaylimits_{m=1}^N a_m |m,A\rangle, \hspace{10mm} |R\rangle = \sum\displaylimits_{m=1}^N b_m |m,B\rangle,
\end{equation}

\noindent where $a_m$ and $b_m$ are given by

\begin{align}
a_m &= a_1 \left( \frac{-v}{w} \right)^{m-1} \hspace{10mm} \label{eqn:a} \\
b_m &= b_N \left( \frac{-v}{w} \right)^{N-m} \hspace{10mm} &\forall m \in \{1, \hdots, N\}, \label{eqn:b}
\end{align}

\noindent and $a_1$ and $b_N$ fix normalisation.

First we calculate the numerator of equation~\ref{eqn:pert}

\begin{equation}
|L\rangle^T \mathcal{J} |L\rangle = 2J\sum\displaylimits_{m=1}^{N-1} a_m a_{m+1},
\end{equation}

\noindent and use equation~\ref{eqn:a} to see that

\begin{equation}
|L\rangle^T \mathcal{J} |L\rangle = -2J\frac{v}{w}a_1^2\sum\displaylimits_{m=1}^{N-1} \left(\frac{v}{w}\right)^{2m-2}.
\end{equation}

Next we calculate the denominator of equation~\ref{eqn:pert}. For the left eigenvalues this is given by

\begin{align}
|L\rangle^T |L\rangle &= \sum\displaylimits_{m=1}^N a_m^2, \\
&= a_1^2\sum\displaylimits_{m=1}^N \left(\frac{v}{m}\right)^{2m-2},
\end{align}

\noindent So the energy of the left edge mode for small next nearest neighbour hopping is given by

\begin{equation}
E_L = -2 J\frac{v}{w} \frac{\sum\displaylimits_{m=1}^{N-1} \left(\frac{v}{m}\right)^{2m-2}}{\sum\displaylimits_{m=1}^N \left(\frac{v}{m}\right)^{2m-2}}
\end{equation}

\noindent In the thermodynamic limit $N \to \infty$, given that $|v| < |w|$ (which is required for edge modes to exist), this converges to 

\begin{equation}
E_L = -2 J\frac{v}{w},
\end{equation}

\noindent as predicted for the Hermitian case by numerical fit~\cite{PerezGonzalez2018}.

For completeness we perform the same calculation for the right edge mode. We have, for the numerator,

\begin{equation}
|R\rangle^T \mathcal{J} |R\rangle = 2J\sum\displaylimits_{m=1}^{N-1} b_m b_{m+1},
\end{equation}

\noindent which when combined with equation~\ref{eqn:b} has

\begin{equation}
|R\rangle^T \mathcal{J} |R\rangle = 2Jb_N^2\sum\displaylimits_{m=1}^{N-1} \left(\frac{-v}{w}\right)^{2N-2m-1}
\end{equation}

\noindent Next we rearrange the sum by making the substitution $j = N - m$, at which point it becomes clear that the calculation is the same as for the left edge modes,

\begin{equation}
|R\rangle^T \mathcal{J} |R\rangle = -2J\frac{v}{w}b_N^2\sum\displaylimits_{m=1}^{N-1} \left(\frac{v}{w}\right)^{2m-2}.
\end{equation}

For the right edge modes the denominator of equation~\ref{eqn:pert}. is given by

\begin{align}
|R\rangle^T |R\rangle &= \sum\displaylimits_{m=1}^N b_m^2, \\
&= b_N^2\sum\displaylimits_{m=1}^{N} \left(\frac{v}{m}\right)^{2N - 2m}, \\
&= b_N^2\sum\displaylimits_{m=1}^N \left(\frac{v}{m}\right)^{2m-2}, \\
\end{align}

\noindent As with the left eigenmode the energy for small next nearest neighbour hopping is given by

\begin{equation}
E_R = -2 J\frac{v}{w} \frac{\sum\displaylimits_{m=1}^{N-1} \left(\frac{v}{m}\right)^{2m-2}}{\sum\displaylimits_{m=1}^N \left(\frac{v}{m}\right)^{2m-2}}
\end{equation}

\noindent Again, in the thermodynamic limit $N \to \infty$, given that $|v| < |w|$ (which is required for edge modes to exist), this converges to 

\begin{equation}
E_R = -2 J\frac{v}{w},
\end{equation}

\noindent so as expected for this symmetric Hamiltonian both edge modes have the same path in complex space as the next nearest neighbour hopping $J$ changes.

\bibliography{Biblio}
\bibliographystyle{unsrt}

\end{document}